# SOLITARY WAVES AND SUPERSONIC REACTION FRONT IN METASTABLE SOLIDS

Hendrik J. Viljoen and Lee L. Lauderback
*Department of Chemical Engineering*
*University of Nebraska-Lincoln*
*Lincoln, NE 68588-0126*

Didier Sornette
*Institute of Geophysics and Planetary Sciences*
 *and Department of Earth and Space Sciences*
*University of California, Los Angeles, CA 90095*
*and Laboratoire de Physique de la Matiere Condensee*
*CNRS UMR6622 and Universite des Sciences, B.P. 70, Parc Valrose*
*06108 Nice Cedex 2, France*

**Abstract**: Motivated by an increasing number of remarkable experimental observations on the role of pressure and shear stress in solid reactions, explosions and detonations, we present a simple toy model that embodies nonlinear elasticity and dispersion as well as chemical or phase transformation. This generalization of the Toda Lattice provides an effective model for the description of the organization during an abrupt transformation in a solid. One of the challenges is to capture both the equilibrium degrees of freedom as well as to quantify the possible role of out-of-equilibrium perturbations. In the Toda Lattice, we verify that the particle velocities converge in distribution towards the Maxwell-Boltzmann distribution, thus allowing us to define a bona-fide temperature. In addition, the balance between nonlinearity and wave dispersion may create solitary waves that act as energy traps. In the presence of reactive chemistry, we show that the trapping of the released chemical energy in solitary waves that are excited by an initial perturbation provides a positive feedback that enhances the reaction rate and leads to supersonic explosion front propagation. These modes of rupture observed in our model may provide a first-order description of ultrafast reactions of heterogeneous mixtures under mechanical loading.

# 1 INTRODUCTION

## 1.1 Experimental motivations

Diffusion transfers of mass or heat usually control front propagation associated with chemical reactions or phase transformations. As a consequence, the velocity of fronts is small and even negligible compared to the sound velocities of the reactants and of the products. Typical solid-solid reactions like $Ta + C \rightarrow TaC$ or solid-liquid reactions like $2Al + Fe_2O_3 \rightarrow Al_2O_3 + 2Fe$, characterized by extremely high activation energies, can react in the combustion mode and these rates are determined by the preheating of reactants by thermal conduction. The combustion front velocity is proportional to $\sqrt{\frac{\kappa}{\tau}}$ where the thermal diffusivity $\kappa = \frac{k}{\rho C_p} \approx \frac{O(10^1)}{O(10^3) \times O(10^3)} \approx O(10^{-6})$ and the characteristic reaction time $\tau = \frac{1}{k_0 e^{-E/RT_{ad}}} \approx O(10^{-2})$. Therefore, the reaction front velocity is of the order of $v \propto \sqrt{\frac{\kappa}{\tau}} \approx O(10^{-2}) m/s$. Thus, diffusive transfer cannot explain events propagating at front velocities much faster than cm/s such as detonations or deflagrations, explosive recrystallization, photo-induced reactions and the high-pressure heterogeneous reactions studied by Bridgman in his pioneering work and later by Enikolopyan.

The ultrafast reaction of heterogeneous mixtures under mechanical loading is particularly intriguing. In 1935, Bridgman reported results of combined hydrostatic pressure and shear for a wide variety of materials [1]. Whilst most substances underwent polymorphic transformation, some reacted rather violently. In contrast to $PbO$, that decomposed quiescently to a thin film of lead, $PbO_2$ detonated and residue of Pb was found afterwards. Reactive mixtures produced even more violent results: stoichiometric mixtures of Cu and S detonated at applied axial load of 2 GPa (even without applied shear), producing $CuS$. More exothermic reactions like $Al/Fe_2O_3$ proceeded in a detonation-like manner, damaging parts of the press - this reaction was initiated at hydrostatic pressure between 1-3 GPa, even without application of shear.

Russian scientists actively continued the work of Bridgman. Enikolopyan and co-workers studied many systems, both endothermic and exothermic in Bridgman anvils and high-pressure extruders [2-10]. They expanded the list of compounds that were originally investigated. Thermite mixtures of $Al$ and $Fe_2O_3$, pressed into discs of thickness 4mm, reacted completely within 100 nanoseconds [7,9] (which, if nucleated from a side would correspond to a velocity as large as 40,000 m/s). The anvil was destroyed and the lack of plastic deformation in its fracture zones points to a detonation. Particles were ejected from cylindrical pre-forms (samples were not radially contained) at velocities up to 2,000 m/s [13]. Reactions were accompanied by the emission of light, high-energy electrons, acoustic emission and gamma radiation [5]. The experiments of Enikolopyan lend further proof to the existence of structural collapse. In order to explain the required level of mixing for these rates, the reactions must have been preceded by decomposition steps which consist of either a phase change (solid to liquid or supercritical fluid) or mechanical disintegration (pseudo-volumetric fracture) or a combination of them. The results for the thermite reaction is particularly intriguing, since the particle sizes are so large ($300 \mu m \leq \phi_{Al} \leq 1,000 \mu m; \phi_{Fe_2O_3} \approx 1,000 \mu m$) that the system would otherwise have difficulty to ignite and react in the normal self-heat-sustained (SHS) mode. To explain such conversion rates on the basis of diffusion and thermal conduction only, the reacting particles would have to be clusters of 5-6 molecules.

Fast decomposition of a metastable phase under strain or through the coupling between chemistry and mechanical strain has also been observed in glassy semiconductors and metals, as well as in a geological context [11, 12]. The "Prince Rupert drops", and more generally, tempered glasses under high strain condition, which explode as a result of a mechanical stimulation [13,14] is a spectacular example, which has remained a mystery for several centuries. Studies of detonation of classical solid explosives (such as heavy metal azides which are compounds bearing the group N$_3$) also reveal the existence of a fast propagation regime with velocities of the order of the sound velocity in solid matrices, before the gas explosion stage [15]. The explosive nature of recrystallization of amorphous materials has been described by Koverda [16]. Hlavacek [17] has observed a clearly distinguishable thermal wave when intensely milled aluminum powder transforms from amorphous (and highly plasticized) state to polycrystalline state. Fortov and co-workers [18] applied high current densities to thin $Nb-Ti$ wires in a cryostat. They have measured propagation velocities for the transformation from the superconduction to the normal conduction phase of $10-12 km/s$.

In a different context, it has been shown that chemical waves propagate at very low temperature [19,20] and at usual temperatures [21] at rather high velocities, due to a coupling between chemistry and mechanical deformations. In the context of cryochemistry of solids, there is evidence of a transition between slow and fast heat-mechano-chemical wave modes of, possibly, gas-less detonation were [22]. This physical phenomenon may be very important, as the fast autowave concept may help to explain the mystery of fast chemical evolution of substances in the universe [23]. It has also been proposed that catastrophic geo-tectonic phenomena, such as earthquakes, may be triggered by gas-less detonation processes of phase transformations in the earth's crust (for example, explosive decay of a metastable glassy state of rocks to a more stable, polycrystalline phase) [24,25]. This hypothesis of phase transformations of rocks induced by a high value of the strain may resolve a number of difficulties with the current purely mechanical theory of earthquakes [24].

The experimental results described above strongly suggest the importance of a coupling between chemistry and mechanical deformations. Batsanov [26] pointed out that, with increased pressure, ionic compounds rearrange valence electron density distributions and the Szegeti charges (i.e., the actual degree of ionization of the atoms in ionic crystals) are reduced until a state of decomposition is reached. Gilman has shown that shear strain changes the symmetry of a molecule or of a solid and is thus effective in stimulating reactions, much more so than isotropic compression [27-29]. The reason is that a shear strain displaces electronic band energies in a different way, thus leading in general to a narrowing of the band gap separating the valence from the conduction bands. When the gap closes, the semi-conductor solid becomes metallic which triggers a strong chemical reactivity. This process belongs to the class of phenomena grouped under the term "mechano-chemistry" [30] and is also at work in the motion of a dislocation by that of kink (leading to plastic deformation). Indeed, the motion of a kink is akin to a local chemical reaction in which an embedded "molecule" is dissociated, and then one of the product atoms joins with an atom from another dissociation to form a new ``molecule" [28]. Gilman has also proposed [31] that intense strain deformation (which therefore lead to bending of atomic bonds) occur in a very narrow zone of atomic scale that can propagate at velocities comparable to or even higher than the velocity of sound in the initial material.

## 1.2 The limits of conventional theory of shock waves and of explosions and the need for out-equilibrium mechano-chemistry

The initiation and propagation of shock waves have been studied for many years and the mainstay of theoretical description is still the macroscopic Rankine-Hugoniot equations, augmented by the Chapman-Jouguet processes [32]. This theory has proven to be remarkably accurate despite its simplicity, but it has certain restrictions and over the years the theory has evolved and it has seen some modifications. For example the Chapman-Jouguet process yields a constant value for the velocity of the reaction front with respect to the unreacted phase. Sano and Miyamoto [33] have developed an unsteady state Rankine-Hugoniot theory and point out the deficiencies of the steady state theories. But arguably the greatest shortcoming of the theory to account for the above mentioned experiments is the lack of detailed description of events that occur on the microscopic level. Erpenbeck [34] combined molecular dynamics (MDS) and Monte Carlo methods to study diatomic exchange reactions and showed qualitative agreement with the CJ theory. An interesting finding of that study is the relatively long time required to reach equilibrium. This raises the question of non-equilibrium reactions, specifically in the setting of a detonation. White and co-workers [35] analyzed a similar system by MDS and their findings also confirmed the validity of the CJ theory only once equilibrium is established. Considerable effort has been focused on excitation and non-equilibrium in molecular crystals, due their importance as secondary explosives. Dlott and Fayer [36] and Kim and Dlott [37] showed that, during the incipient phase after shock loading, bulk phonon modes are excited first and then energy is transferred to intra-molecular modes through doorway modes that are most conducive to the transfer. In the case of naphtalene, equilibrium is only reached 200ps after a 40kbar impact. Coffey [38,39] studied the interaction of molecules with compressive waves and developed a model for multi-phonon excitation of intra-molecular vibrational modes. These models have proven invaluable in the interpretation of experimental results under conditions where thermodynamics equilibrium is reached at the microscopic scale. However, as summarized in section 1.1, there are now an increasing number of experiments which cannot be understood within the conventional framework.
.

## 1.3 The two fundamental mechanisms

Two mechanisms have recently been identified as potential candidates for explaining very fast phase transformation and explosions in solids: 1) the products are mechanically stronger than the reactants leading to mechanical shock waves [25, 40, 41]; 2) the released energy is sufficiently large and its release rate is so fast that its transfer to microscopic degrees of freedom literally boosts and propels atoms to collide against each other leading to supersonic chemical fronts [29, 31, 25].

Courant and Friedrichs [31] have studied wave propagation of finite amplitude in elastic-plastic materials and pointed out that shocks are not possible when the stress-strain characteristics of plastic material is of the weakening type. Sornette [25] adapted their one-dimensional formalism to study the opposite case in which the products are more elastically rigid than the reactants, and the density is smaller for the products than for the reactions, leading to a larger sound velocity for the products. Under these rather special conditions, a shock develops which propagates at a velocity intermediate between that the acoustic wave velocities of the reactions and products. Consider a bar of material deformed uniformly with an initial strain everywhere along it. Suppose that a localized perturbation or inhomogeneity produces a local deformation larger than the initial strain at the left boundary of the bar. Qualitatively, the density perturbation will start to advance to the right in the product phase. Since the velocity is larger in the product phase, the largest deformations propagate the fastest. An initial smooth disturbance will progressively steepen and a shock will eventually form. The shock is fundamentally due to the rigidifying transformation from reactants to products. The importance of this condition was independently recognized by Pumir and Barelko [40,41], using a slightly different formulation. Their framework coupling the elastic wave equation to a reaction-diffusion equation allowed them to reveal the existence of supersonic modes of deformations with the existence of a critical strain necessary to ignite gasless detonation by local perturbations.

The second mechanism discussed by Gilman [29, 31] is illustrated in figure 1 (see also Ref. [25]). A one-dimensional chain is made of atoms of mass m linked to each other by energetic links of spring constant k which, when stressed beyond a limit, rupture by releasing a burst of energy $\Delta g$ converted into kinetic energy transmitted to the atoms. Initially the chain of atoms is immobile. Suppose that the first atom on the left is suddenly brought to a position that entails the rupture of the first bond. This rupture releases the energy Δg which is converted into kinetic energies of the atom fragment that is expelled to the left and of the next atom to the right which becomes the new left-boundary of the chain. Now, due to the impulsive boost $(\Delta g/m)^{1/2}$ that the boundary atom received, it will eventually stress the bond linking it to the next atom towards its rupture threshold. When this occurs, it is expelled by the energy that is released and the next atom forming the new boundary is itself boosted suddenly by the amount $(\Delta g/m)^{1/2}$. It is then clear that this leads to a shock propagating at a velocity larger than the sound velocity equal to $c=(k/m)^{1/2}$ in the long wavelength limit, since the atoms are receiving boosts that accelerate their motion faster than what would be the propagation by the springs with the usual acoustic wave velocity. Taking the continuous limit, Gilman [26, 28] proposes that the resulting supersonic shock velocity U is given by the Einstein formula $U^2 = c^2 + \Delta g$.

**1.4 Our modeling strategy using the reactive Toda lattice**

The understanding of the conditions under which these two mechanisms will be active in real materials is at a rudimentary stage, not to speak of their possible interplay. We need models that allow us to explore the relationship between non-equilibrium processes at the microscopic scale, the chemical reactions and the possible development of supersonic shocks and explosions. In this goal, we propose to use a mesoscopic approach describing how the microscopic processes self-organize into dynamic macroscopic structures of patterns and waves. The fundamental problem, and therefore the challenge, in any mesoscopic model lies in the compromise between scale (macroscopic limit) and detail (microscopic limit). The compromise proposed here is based on the Toda lattice [42], which is a one-dimensional system of entities interacting through an interaction potential limited to nearest neighbors. Using a coarse-grained discrete lattice allows us to perform simulations on large systems at times large compared to all the relevant times scales so that the characteristics of the self-organization behavior can be studied in details.

The advantage of the Toda lattice are multifold. First, it recovers linear elasticity and the non-dispersive acoustic waves (phonons) in the large wavelength and small amplitude limits. Its non-linear potential at large deformation leads to mode-coupling which, as we shall briefly discuss, gives naturally the Maxwell-Boltzmann distribution of energies. Thus, with a purely deterministic dynamics (Newton's equations) on a minimal model, we have a basic thermodynamics that we can then enrich to study chemical reactions coupled with mechanical strain. The Toda lattice has dispersion at non-vanishing frequencies, which is adequate to capture the effect of the microscopic atomic structure as well as any possible mesoscopic organization. The Toda lattice has been extremely well studied in the literature as a remarkably simple system exhibiting stable localized collective excitations, called solitons. Solitons are a particular set of a more general class of solitary waves or moving discrete breathers found in many lattice systems with both dispersion and anharmonic interactions [43]: the dispersion tends to disperse the modes while the non-linearity tends to concentrate them. The resulting behavior is a localized coherent nonlinear wave. We stress that our choice of the Toda potential does not restrict our conclusions as similar solitary waves can be observed for a large class on non-linear potentials. In particular, we have checked that Taylor expansions of the exponential terms in the Toda potential truncated at different orders do not change our results at the qualitative level. We think that the result reported below are robust to a generalization of the Toda potential. Indeed, recently, the existence of solitons has been established in anharmonic lattices for a large class of interatomic potentials [48,49].

We modify the classic Toda model to include metastable states and the possibility for a phase transformation or chemical reaction. Specifically, beyond a certain strain threshold, the potential felt by the particles of the lattice is modified to represent a change of phase from the reactants (initial lattice) to the products. In contrast to regular dispersive waves, solitons act as energy traps because they create a dynamic state where the local energy flux points in the direction of wave propagation. As a consequence, the potential energy, which is released during the reaction as kinetic energy, can be trapped within the soliton, enhancing its localization and its velocity. This process occurs out-of-equilibrium, i.e., without equilibration with other degrees of freedom and can thus focus energy to extremely high levels. In our knowledge, a Toda lattice with chemical reaction has not been studied before.

More generally, the Toda lattice should also shed light on several issues in ultra-fast solid phase reactions that are currently not understood. When the shock wave travels through a particle, either fracture or melting must occur, because the high reaction rates can only be explained by the generation of large surface areas.

Toda [42] mentions that the Toda lattice has a property called "chopping phenomenon". When a soliton, which is a compressive pulse, is reflected at a free boundary a temporary tensile pulse develops that could be causal in spalling or fracture near the free boundary. Since the tensile soliton is unstable it degenerates into a series of pulses and there is a rapid transfer of energy from the soliton to thermal vibrations of the lattice.

The paper is organized as follows. In section 2, the classic Toda lattice (without reaction) is introduced and the lattice equilibrium is investigated. Local departures from equilibrium are demonstrated for solitons in inert lattices (with no metastable states) and the problem of non-equilibrium is discussed. In section 3, the reactive lattice is defined and studied. The model is easily generalized to an axi-symmetrical 2-D geometry. Some of the interesting results involve the boundary conditions and wave inversions lead to hot spots and spalling. Section 4 is a discussion of the results. The connection is made between this model and photo-induced first order phase transition models.

## 2 THE CLASSIC TODA LATTICE (TL) MODEL
### 2.1 Definition

The Toda lattice is a model of a one-dimensional chain of atoms. Consider a one-dimensional lattice consisting of N particles. Each particle is described by a point of mass $m_n$ that only interacts with neighboring masses. Hence, heterogeneities, discontinuities, pores and perfectly isotropic states are all defined by the pair-wise values of the interaction potential and masses. The displacement of the $n^{th}$ mass from its equilibrium position is $y_n(t)$. The relative displacement is defined as $r_n = y_{n+1} - y_n$. The lattice motion is described by the following canonical equations:

$$m_n \frac{d^2 r_n}{dt^2} = \phi'(r_{n+1}) - 2\phi'(r_n) + \phi'(r_{n-1}) \tag{1}$$

where $\phi'$ denotes the first derivative of the potential function with respect to the relative displacement. In his search for an integrable lattice that also exhibits realistic mechanical behavior, Toda used a recursive formulae to find periodic and single soliton solutions. As a result, the potential function

$$\phi(r) = \frac{b}{a} e^{-ar} + br, \quad (ab > 0) \tag{2}$$

was proposed. Application of this choice in eq. (1) gives the Toda lattice (TL). Note that the integrability of the Toda lattice makes this system (1) special because it exhibits by construction an infinite number of invariants [42,43]. The TL equations can also be written in terms of the displacement:

$$m\frac{d^2 y_n}{dt^2} = b[e^{-a(y_n - y_{n-1})} - e^{-a(y_{n+1} - y_n)}]. \tag{3}$$

While the TL model is 1D and mesoscopic in nature, we still want to use parameters that are realistic. As an illustration, we use the properties of *Al* to determine the parameters in eq.(3). If $a = 1/\lambda$, where $\lambda$ is the athermal lattice constant = $4.5 \overset{o}{\text{A}}$ and the cold longitudinal sound velocity is $c_0 = 6,420 m/s$, it follows that $b = 4.1 \times 10^{-9} N$. The model is written in non-dimensional form: $t \to tc_0/\lambda$ and $y \to y/\lambda$. Later, we will look at the effect of chemistry in a simplistic manner by changing the parameters of the potential function (more details in section 2.3). If we call $a'$ and $b'$ some new parameter values due to the chemical reactions that will be introduced in section 2.3, we will define the dimensionless parameters $\alpha = a'/a$, $\beta = b'/b$. Thus, $\alpha = \beta = 1$ are the values used to model the standard (non-reactive) lattice. The non-dimensional TL equation becomes

$$\frac{d^2 y_n}{dt^2} = \beta[e^{-\alpha(y_n - y_{n-1})} - e^{-\alpha(y_{n+1} - y_n)}]. \tag{4}$$

**2.2 Energy partitioning in the Toda Lattice**
**2.2.1 The Maxwell-Boltzmann distribution**

If the TL is a fair description of a chain of atoms, one would expect that in the absence of external forcing, the lattice should approach a thermodynamic equilibrium state. Oscillators that are initially more excited should exchange energy with neighbors so that in the end the Maxwell-Boltzmann distribution is recovered. It turns out that the TL model is not ergodic (due to the integrability property), but that does not exclude energy sharing. This state is described by the Maxwell-Boltzmann distribution for one translational degree of freedom. Indeed, since the total energy of the system is conserved (in the absence of dissipation), this corresponds to the micro-canonical ensemble and the relevant observables are the energy of the microscopic degrees of freedom.

For the sake of brevity, let us define the instantaneous velocity of the $n^{th}$ oscillator as $c$ and let us decompose it as the sum of a drift or wave component $w$ and of a random fluctuating component $v$. When the oscillations are resolved on the time scale associated with the Debye frequency (which is basically the period of a thermal vibration), the question arises: how do we discriminate between motion associated with thermal fluctuation and motion associated with an event like a wave that propagates through the lattice? The oscillator does not discriminate, it only responds to the immediate force, be it of thermal or mechanical (wave) origin. The numerical results of Fig.2 pertains to a situation where no external force was applied, in other words, we did not have a drift component $w$ and the velocity of each oscillator was taken to be $v$. The stochasticity was introduced through the initial conditions of the oscillators; the initial conditions were zero velocity for all oscillators and random displacement between [-W, +W] where W is an input parameter for the system. The larger W, the larger the input energy. The Toda Lattice was then integrated for random initial displacement and zero velocity and the stochastic component of the velocity was equal to the total velocity.

If the chain is in equilibrium, it has a Maxwell-Boltzmann distribution parameterized by $T$. This means that the probability to find a state of energy E is $e^{-E/kT}$, where k is the Boltzmann constant (whose sole use is to convert a temperature scale into an energy scale) and T is the (temperature) parameter quantified the degree of excitation or disorder of the vibrations within the lattice. Using the form (2) for the potential energy, the Maxwell-Boltzmann distribution reads

$$f = A e^{-2\gamma[\Phi + \frac{1}{2}v^2]/T} \quad \text{with } \gamma = \frac{mc_0^2}{2k} \text{ (units K)}, \tag{5}$$

where we have used the expression of the total energy of a given oscillator (a spring + mass) as the sum of the potential energy stored in the spring and the kinetic energy associated with the stochastic component of the mass velocity.

The dimensionless potential function is defined as

$$\Phi = \frac{\beta}{\alpha}[e^{-\alpha(y_{n+1}-y_n)} - 1] + \beta[y_{n+1} - y_n]. \tag{6}$$

This distribution (5) should be calculated by counting the number of oscillators in a given energy bin, For this, we integrate the non-dimensional TL equation (4) (with $\alpha = \beta = 1$ since chemistry is not considered here) and construct the histogram of the energy of individual oscillators using the cumulative statistics over all the elements in the lattice and over time. Assigning random initial displacements of the lattice points uniformly distributed in $[-W, W]$, the density distribution of the oscillators over energy space has the form shown in Fig.2 where it is compared with the Maxwell-Boltzmann distribution. W is thus the characteristic scale of the energy put initially inside the TL.Fig.2 has been constructed with the choice $W = 0.081$ which is best matched by the Maxwell-Boltzmann distribution at the temperature $T = 297K$. We verify by launching several runs that T is proportional to W, indicating that higher input energy results in higher temperature. This is an important test of our model. In addition, we kept the boundaries free in these numerical experiments and observed thermal expansion. It is actually quite interesting that the linear expansion that occurs compares quite favorably with the linear (volumetric/3) expansion coefficient of aluminum (recall that the a and b values of the potential correspond toaluminum).

To further test for the relevance of the Maxwell-Boltzmann distribution, we have also constructed the distribution over a single oscillator (for instance the $50^{th}$ oscillator) and by summing the statistics over a time interval Δt. As Δt increases, we verify that the energy distribution of a single oscillator is also well-described by the Maxwell-Boltzmann distribution. This result implies that the stochastic component of a typical oscillator velocity added to its potential exhibit an approximate ergodicity property.

**2.2.1 Non-equilibrium configurations**

The quest for equipartition of energy and for the Maxwell-Boltzmann statistics from nonlinear dynamics was first initiated by Fermi, Pasta and Ulam who failed [44]. More recent works (see for instance [45]) have shown the subtlety of this problem. In their pioneering work, Fermi, Pasta, and Ulam revealed that even in strongly nonlinear one-dimensional classical lattices, recurrences of the initial state prevented the equipartition of energy and consequent thermalization. The questions following from this study involve the interrelations between equipartition of energy (Is there equipartition? In which modes?), local thermal equilibrium (Does the system reach a well-defined temperature locally? If so, what is it?), and transport of energy/heat (Does the system obey Fourier's heat law? If not, what is the nature of the abnormal transport?). The surprising result of Fermi, Pasta and Ulam has now been understood: under general conditions for classical many-body lattice Hamiltonians in one dimension, it has been shown that total momentum conservation implies anomalous transport in the sense of the divergence of the Kubo expression for the coefficient of thermal conductivity [46]. The anomalous transport is thus a specific feature of one-dimensional systems. Thus, our verification of an approximate Maxwell-Boltzmann distribution does not prevent the existence of anomalous transport or propagation properties as we discuss in the sequel.

It is indeed possible to perturb a small region of the lattice in such a way that these nodes depart from the equilibrium distribution and this perturbation propagates with conservation of form and of energy. If the chain contains many nodes, the non-equilibrium state of the small number of nodes perturbed in this coherent mode (namely forming a soliton) will not significantly alter the overall distribution. The transit time of a soliton over a given oscillator is very small compared to the time scale over which equilibrium is achieved at the scale of a single oscillator. The transit of a soliton is thus fundamentally a non-equilibrium process. To recognize this fact is essential for our investigation of the coupling with a chemical reaction performed below.

Indeed, when a shock wave propagates through the lattice, the distribution of the energies of the oscillators in the shock zone is perturbed away from its Boltzmann distribution. Depending on the magnitude of the shock (ranging from a sound wave to a detonation), the degree of deviation from equilibrium could vary between insignificant to complete. Quantifying this degree of deviation from equilibrium constitutes one of the major dilemmas of shock theory: within the shock zone, non-equilibrium could exist, temperature could become meaningless and a macroscopic description of chemistry with Arrhenius kinetics becomes nonsensical. Our mesoscopic approach allows us to investigate precisely this regime and the interplay between the equilibrated degrees of freedom and the out-of-equilibrium impulses.

To study the effect of an external perturbation, consider a force that is applied at the first node:

$$F_L = M \sin(\frac{\pi t}{t_I}), \ t \leq t_I \text{ and } F_L = 0, \text{for } t \geq t_I. \tag{7}$$

This impulse corresponds to a single arch starting from and returning to 0. The motion of the first node is then described by

$$\frac{d^2 y_0}{dt^2} = \beta[1 - e^{-\alpha(y_1 - y_0)}] + F_L. \tag{8}$$

The last node could be either fixed or unbounded. When the force is small, the deviation from equilibrium is slight. Application of a stronger force at the first node will cause the creation of a solitary wave leading to a strong deviation from equilibrium as illustrated in a phase diagram for the $70^{th}$ oscillator shown in Fig.3. The conditions and parameter values for this example are listed in Table I as case 1. The oscillator has been in equilibrium before the first arrival of the shock wave as can be seen from the chaotic trajectory in phase space shown in Fig. 3. When this oscillator is shaken by the passage of the shock, it rapidly moves away from its equilibrium state to a novel state, a new position in phase space where it settled chaotically and equilibrates over time. Due to the finiteness of the system used in the simulations, the shock will reach the boundary of the lattice and be reflected at the last node (we use a fixed boundary condition at the extremity $n = 200$ and a free boundary at the other extremity). When the $70^{th}$ oscillator is again shaken by the reflected solitary wave, it undergoes a second translation. Between these strong perturbations, the oscillators can reach equilibrium provided the time intervals between the solitary wave perturbations are much longer than a vibrational period. In this example, the $70^{th}$ oscillator started with a temperature of $297K$ before the solitary wave arrived the first time and finished with the temperature of $328K$ after the solitary wave was completely dissipated and final equilibrium was restored globally in the lattice. Before this final equilibrium was reached, the oscillator has been displaced several times by the weakening multiply reflected solitary wave.

In the next example (case 2), the duration $t_f$ of the applied force is extended ten times such that it becomes slightly larger than twice the dimensionless natural period $\frac{2\pi}{\sqrt{2\alpha\beta}}$ of oscillation of the oscillators, obtained by expanding eq. (4) to linear order and neglecting dispersion (Einstein approximation. The strain (relative displacement) profiles along the lattice of cases 1 and 2 are shown in Fig. 4 at $t = 100$. There is a distinct difference between the two profiles. In case 1, the solitary wave travels as a single perturbation through the medium with velocity of Mach 1. In case 2, the initial perturbation has split into two solitary waves. The leading one has velocity of Mach 1.08 and the second wave travels at Mach 1.02. As a consequence, as time progresses, the distance between the two waves of case 2 increases. The solitary waves conserve their form as they pass through each other, consistent with the original studies of Zabusky and Kruskal.[47], defining them as "solitons" At the free boundary, the solitons are destabilized and their energy is transferred to the lattice. The energy near the free boundary thus increases after the soliton has been reflected and a 'hot spot' is generated.

The system was integrated until equilibrium at which the temperature finally reached system after equilibrium is $517K$ in case 2 compared to $328K$ in case 1. The higher temperature of case 2 is expected, since considerable more energy has been injected into the system. An interesting observation is the creation of multiple waves whose number is given by the integer of $t_f/5$. For $t_f$ less than 5, no solitary structure was observed, only strong concentrated acoustic waves propagating at Mach 1 and dissipating within the lattice at they propagate. When the (dimensionless) impact time are 5, 10 and 20, one, two and four solitons are formed respectively. Another interesting result is the effect of impact strength. When $M$ decreases, (regardless of impact time), the perturbation does not split into separate solitons. Instead, it travels as a single perturbation at Mach 1 through the lattice. Although this critical value of $M$ has not been determined precisely, it is clear that a minimum strength of impact is required to create the soliton. When the impact strength is small, the system is adequately described by a linear system (linearization of the exponential terms) and the velocity of any wave is determined by material properties alone. When the impact strength increases, the nonlinear interaction becomes prominent and the wave speed is no longer determined by material properties alone, but also by the magnitude of the impact. This reflects the fundamental property of solitons and more generally of solitary waves to result from the competition between dispersion (which tends to have waves at different frequencies propagate at different velocities) and nonlinearity (which tends to concentrate waves into sharp shocks).

The stability of solitons and the instability of dark solitons are thoroughly investigated in the literature (cf. [42]). Solitons keep energy focused in a small region of the chain and the energy does not disperse to the rest of the chain - a curiously stable, albeit non-equilibrium state. This is a clear indication that energy transfer in the soliton must occur exclusively in the direction of propagation. The energy flux is only directed in the direction of propagation. This property is essential to understand our results reported below of the coupling between the mechanical deformations associated with solitary waves and chemical reactions occurring in the lattice. Indeed, if chemical energy is released in the soliton and if it retains its coherent non-dispersive localized structure, it is possible to focus energy within this solitary wave to very high levels.

**3 CHEMICAL REACTION ON THE TODA LATTICE**

The term `chemical energy` is used in a loose sense. We envision the following experimental situation in which a mechanical system can suddenly undergo a phase transformation or a chemical reaction when its local mechanical deformation reaches a threshold. This can occur, for instance, according the mechanisms of Batsanov [26] and Gilman [27-29] in which the distortion of the lattice by shear strain moves the electronic bands, leading eventually to a closing of the band gap (metallization) and therefore to a sudden strong chemical reactivity. In our system, we shall account for the existence of a phase transformation or of a chemical reaction by introducing a potential energy source when the relative displacement (strain) $r_n=y_{n+1}-y_n$ at some point in the lattice reaches a critical threshold $r_c$. Thus, when two nodes are sufficiently compressed so that the relative displacement $r_n$ becomes smaller than the threshold $r$, we assume that the initial system (reactants) is transformed into a new phase (products) characterized by different parameters for the potential function defined in the right-hand-side of eq.(4), as shown in Table 1. Note that the transformation only occurs and the `chemical energy' is released only when the material is in the compressed state. In this description, the chemical energy could refer to the energy difference between phases, to an amorphous-crystalline transformation or a chemical transformation. Chemical reaction between different species necessarily involves the issue of mixing, which is not addressed in the present study.

Some complication may occur as a chemical reaction brings additional time scales into the system. Here, we assume that the time scale for the release of chemical energy is much shorter than the time scales involved in the wave propagations. It can thus be considered to be a discontinuous step in the Hamiltonian of the system.

To investigate the interplay between mechanical waves and chemical energy release at a phenomenological level, the interaction potential is irreversibly changed into the following expression

$$\Phi_f = \frac{\beta_f}{\alpha_f}[e^{-\alpha_f(y_{n+1}-y_n+\frac{\lambda-\lambda_f}{\lambda})} - 1] + \beta_f[y_{n+1}-y_n], \text{ when } r_n \text{ becomes smaller than } r_c. \quad (9)$$

The lattice constant $\lambda_f$ of the product state could be smaller or larger than $\lambda$. Three different cases are investigated, the first case (case 3 in table I) describes a product that is more dense (i.e. $\lambda_f < \lambda$), the second case (case 4 in table I) describes a material with larger lattice constant ($\lambda_f > \lambda$) and smaller sound velocity and the third case (case 5 in table I) describes a product with larger lattice constant and larger sound velocity than the initial material. The parameters are listed in Table I. (The interaction potentials before and after the reaction are made continuous at $r = r_c$ by adding a constant to $\Phi_f$).

*Case 3(cf. Table I):* $\lambda_f = 0.9\lambda$

Never does a soliton form in this case when chemistry is active. The initial perturbation is strong enough to initiate reaction. The wave propagates supersonically through the medium, but it appears random, like a thermal wave. However, the leading part of the wave is always compressive. But this compressive state weakens and about midway through the lattice, the conversion criterion is not met. The conversion halts and an ordinary acoustic wave continues to propagate through the remainder of the lattice with no further conversion. This is surprising as solitons do form in absence of chemistry. Indeed, with the values of $(\alpha_f, \beta_f) = (1.5;3.0)$ – but no reaction (by setting the initial and final values equal to $(\alpha_f, \beta_f)$ (then $r_c$ becomes irrelevant) and setting $\lambda_f = \lambda$ – we found solitons for perturbations M and $t_I$ values as listed in Table I. For $(\alpha_f, \beta_f) = (0.9;0.8)$ and $(\alpha_f, \beta_f) = (1.1;1.5)$, we also found solitons. These solitons differed in the time to develop and their relative positions with respect to each other. In summary, we can state that for all ``non-reactive'' cases and for $(\alpha_f, \beta_f)$ as listed in the table, solitons exist. Comparing this result with the reactive systems, we can conclude that the instantaneous transformation of the lattice parameters by chemical reaction can destroy the soliton structure.

The relative displacement is shown in Fig.5a-b at t=34 and t=110. Fig. 5a shows a typical profile of the displacement field in the regime when the conversion is still active, i.e., the criterion for phase transformation is met during the wave propagation. Fig. 5b shows a profile when the conversion has ceased to be active as the energy is not sufficiently concentrated to reach the threshold of conversion. The lattice position where the conversion ceases depends on the initial impact energy. For example, when the simulation is repeated for M=0.4 the reaction stops earlier.

### *Case 4 (cf. Table I):* $\lambda_f = 1.1\lambda$ **with a smaller sound velocity**

A soliton is not observed for this case and the situation is very similar to the previous case. The strain profile in Fig. 6 shows an expansion of the lattice behind the wave front. Therefore, the conversion is interrupted soon after onset because the compression is not strong enough. It is clear that the energy that is released in the reaction does not contribute towards sustaining and strengthening of the compression in the leading part of the wave. Beyond the point where the criterion is no longer met, the wave propagates as an acoustic wave through the system.

### *Case 5 (cf. Table I):* $\lambda_f = 1.1\lambda$ **with a larger sound velocity**

A solitary wave is formed in the reactive lattice and it appears to have all the properties of a soliton. Of course, it is not a genuine soliton since its front separates two different phases, with the products left behind it and the reactants in front of it. All the chemical potential energy that is released during the reaction remains trapped in the solitary wave, which travels supersonically. Both the compression and particle velocities in the wave increase, thus the released chemical energy is stored as kinetic and elastic potential energy.

Note that, after its reflection from the fixed boundary at $n = 200$, the solitary wave has "consumed" all the reactions and the system is made entirely of the new phase or products. In this new phase, it becomes a bona fide soliton, traveling unaffected through the other waves in the product material. No further chemical transformations occur and the soliton now travels with constant speed.

# 4 DISCUSSION

We have presented a numerical study that suggests that solitary waves may play an important role in explosive supersonic reactions and conversions. For this, we started from the classic Toda Lattice (TD), extending it beyond its use as a phenomenological model for processes occurring at the atomic level. Since most of experimental observations are performed at the meso- and macro-scopic levels, we use the TL as a simple toy model which embodies the minimal number of essential ingredients of the problem (non-linear elasticity+dispersion) to explore the different possible regimes. We stress that the phenomenology documented and summarized in this paper for the TL is qualitatively in agreement with that obtained from a continuous description based on the Boussinesq equation that can be further transformed into the Korteweg de Vries equation. We have then extended the TL to include chemistry in order to describe a reactive material in non-equilibrium situation, with applications to explosions and detonations as well as very fast phase conversion.

The TL equations are deterministic but the oscillators clearly adopt the Boltzmann distribution, thus providing the correct analog of a thermodynamics. The equilibrium state has been calculated by the probability of finding a given velocity state space both using an ensemble statistics over a large number of oscillators at a fixed moment in time and by cumulative the statistics over a long period of time for a single oscillator (effective ergodicity property). We have clearly documented that the oscillators approach the Boltzmann distribution at long times, that the state variables are defined and they can be matched with the continuum model.

When perturbations as in the shock wave occur on time scales too short to allow the oscillators to reach the Boltzmann distribution, one needs a model that takes suitably into account both equilibrium and out-of-equilibrum states and their interactions. In this sense, the TL is ideally suited. We have in addition build-in chemistry by modifying the Hamiltonian at reactant interfaces to a double well potential: when a compressive stress threshold is reached, reactants combine irreversibly to give products that are modeled by a different interaction potential. Chemical energy is released when the mechanical vibrations resulting from both the possible existence of a solitary wave and of the rest of the vibrational background forces a node to reach the conversion threshold and thus crosses the energy barrier.

We have observed that the released chemical energy initially contributes to the kinetic energy of the oscillator. Due to friction (as in the Langevin equation), this energy tends to be transformed over time into vibrational energy or heat in the transition zone behind the shock wave. If solitary waves can be nucleated by the initial conditions, we have shown that they act as traps of energy: in the presence of on-going chemical energy release, the energy piles up within the trapping solitary wave, providing a positive feedback which enhances the strength and velocity of the solitary wave. The trapping property of solitons and of solitary waves result from the fact that the energy flux within them points forwards in the wake. In contrast, the energy flux that points backwards in the wake of ordinary dispersive waves resulting into a negative feedback: chemical reactions thus tend to dampen out and stop in absence of solitary waves.

When chemistry is included in the lattice, a non-steady state solitary wave is only observed for the case where the product expands with respect to the reactant and the system has higher sound velocity. It is not certain that this is a universal result as there seems to be subtle interply between the formation of solitary waves and the reaction chemistry that can either be stored and enhance the strength of the soliton or on the contrary destroy it. The situation thus seems more complex and rich that obtained previously in [25] and [40,41]. As long as reactants are available, it converts the material and traps the chemical energy as elastic potential energy and the balance as kinetic energy. Within the spatial zone in which the chemical reaction actually takes place, the lattice is highly compressed and particles have large particle velocities. These velocities belong to the drift component and not the random (or thermal) component of the total velocity. The chemical transformation is athermal as the system is far from equilibrium. Furthermore, the chemical energy is not directly released as thermal energy. No energy spills over into thermal energy as long as free boundaries or internal defects are absent. At defects and free boundaries, conversion to thermal energy occurs and eventually leads to thermal equilibrium at long times.

When the product constitutes a contracted state with respect to the reactant, a supersonic wave is observed directly after impact. But it is not self-sustaining and disappears, leaving only an acoustic wave. A similar observation is made for a product with larger lattice constant and lower sound velocity. There results seem to hold true in all the simulations we have performed until now which explored a significant part but nevertheless non-exhaustive fraction of the phase diagram.

It is an ideal model to study non-equilibrium chemistry and lattice motions. The limitations of this model must also be recognized, especially its one-dimensionality and interaction is limited to nearest neighbors. Although 3-D MDS are computationally costly, they should be used to verify the TL findings. This study is underway.

**TABLE I**

M is the maximum amplitude of the force impulse applied at one extremity of the chain. $t_I$ is the duration of the forcing applied at the boundary which creates a train of solitary waves disturbing the equilibrium. The parameter $\lambda$ (resp. $\lambda_f$) is the lattice constant of the reactants (resp. the products if there is a transformation). When both are equal (i.e., $\lambda/\lambda_f=1$), there is no chemical reaction and the simulations corresponds to the classic Toda lattice. The parameters $\alpha_f$ and $\beta_f$ are the ratio of the two parameters defining the Toda potential taken for the reactions over those taken for the products. The values $\alpha_f = \beta_f = 1$ corresponds to the absence of any chemistry, i.e. to the standard Toda lattice. $r_c$ is the threshold for compression at which the phase transformation or chemical reaction is triggered.

| Case No. | M | $t_I$ | $\lambda_f / \lambda$ | $r_c$ | $\alpha_f$ | $\beta_f$ |
|---|---|---|---|---|---|---|
| 1: no reaction | 0.5 | 1. | 1. | _ | 1. | 1. |
| 2: no reaction | 0.5 | 10. | 1. | _ | 1. | 1. |
| 3 | 0.5 | 10. | 0.9 | -0.3 | 1.5 | 3.0 |
| 4 | 0.5 | 10. | 1.1 | -0.2 | 0.9 | 0.8 |
| 5 | 0.5 | 10. | 1.1 | -0.2 | 1.1 | 1.5 |

**Figure 1:** Cartoon of a one-dimensional chain made of blocks linked to each other by energetic links which, when stressed beyond a given deformation threshold, rupture by releasing a burst of energy converted into kinetic energy transmitted to the blocks. The figure shows two successive bond ruptures that lead to velocity boosts to the ejected fragments on the left and to the boundary blocks.

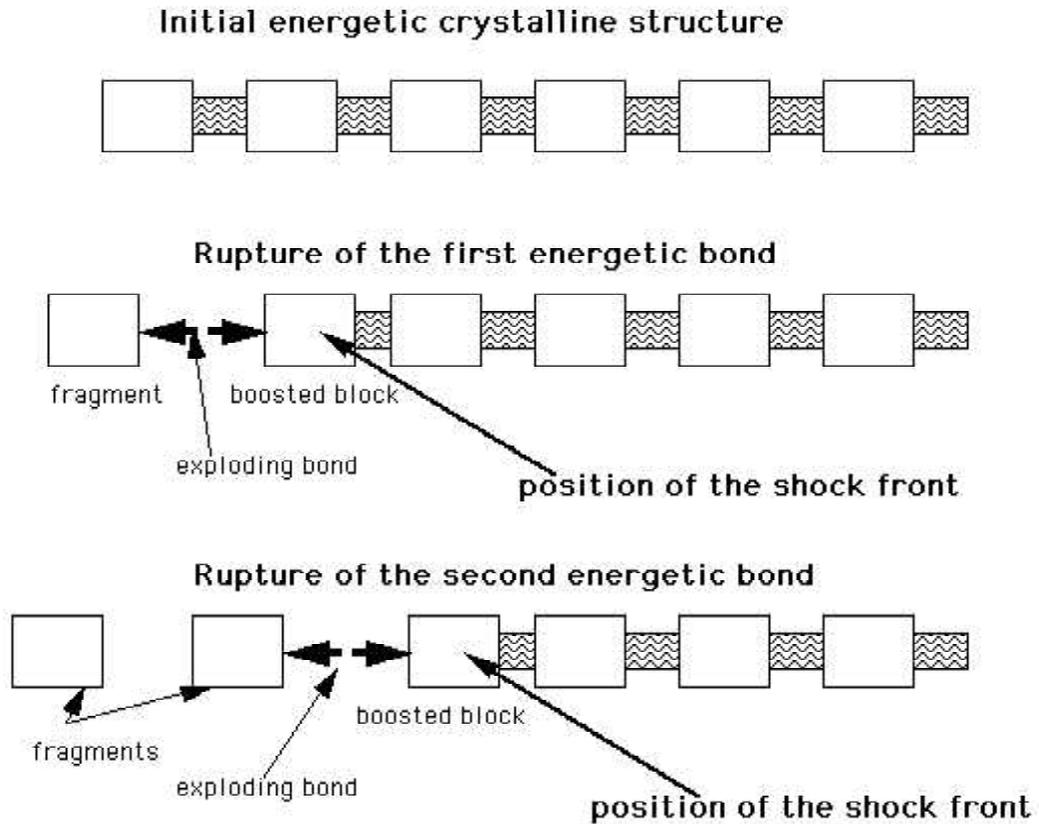

**Figure 2:** Maxwell-Boltzmann distribution (- -) and TL distribution for a chain of 201 oscillators integrated over a total reduced time equal to 500.

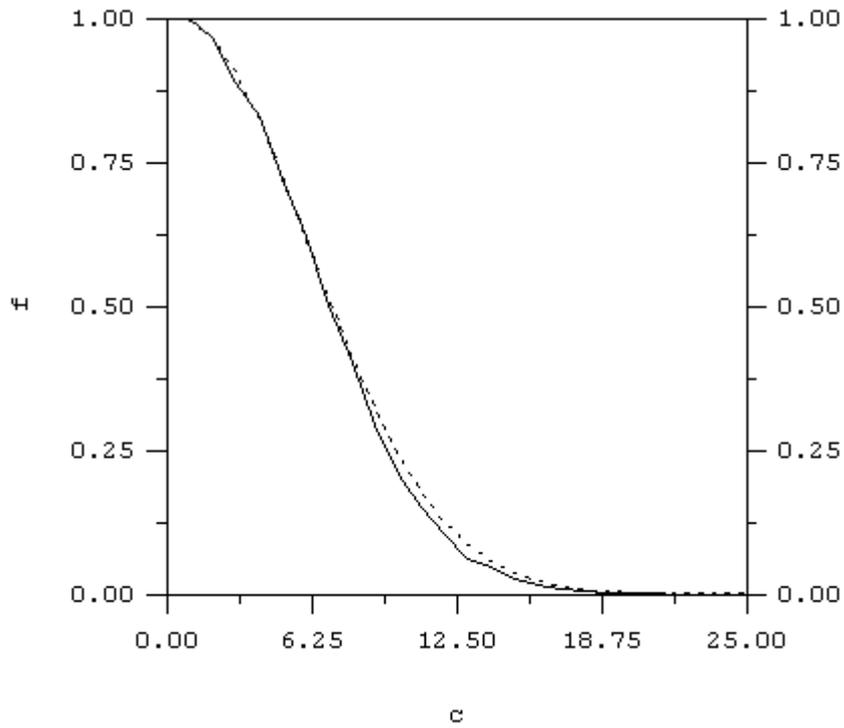

**Figure 3:** Phase diagram for the $70^{th}$ oscillator showing velocity as a function of displacement from the equilibrium position before (left chaotic region) and after (right chaotic region) a strong impact of the oscillator by a solitary wave.

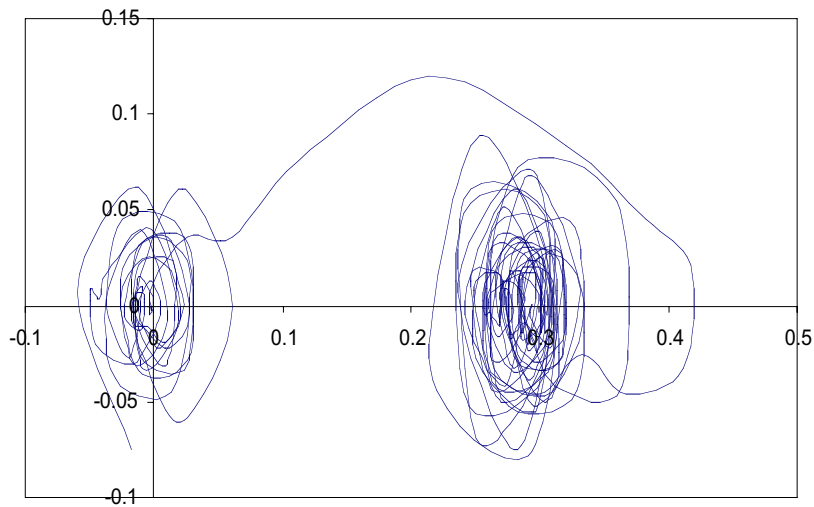

**Figure 4:** Strain profiles at time t=100 of case 1 (a) and case 2 (b) showing respectively 1 and 2 solitons propagating over a noisy "thermally" equilibrated background.

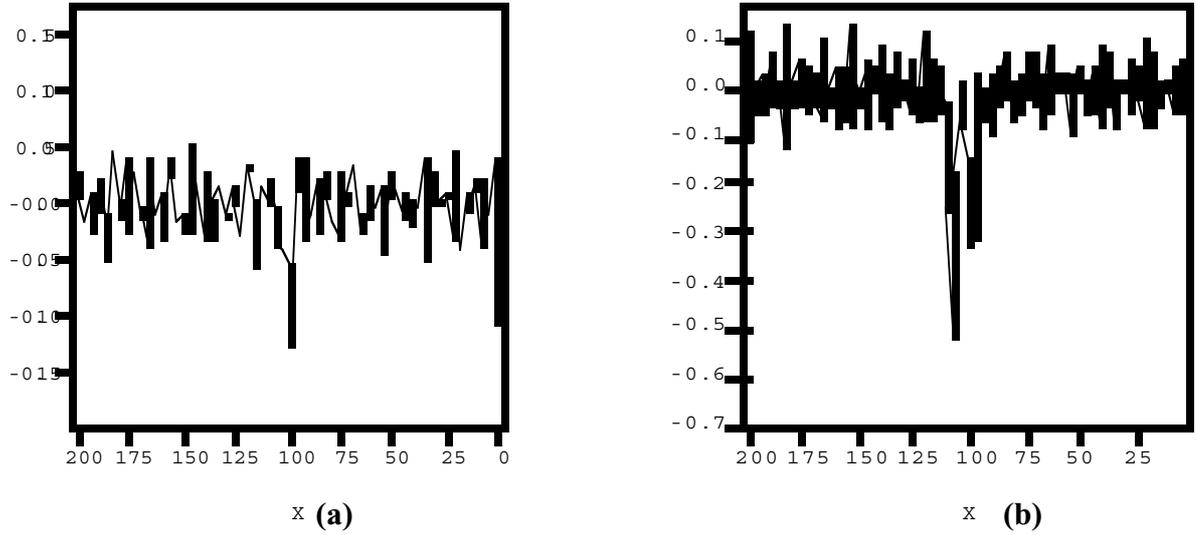

x **(a)**     x **(b)**

**Figure 5:** Strain profiles for case3, at a) t=34 and b) t=110.

**(a)**     **(b)**

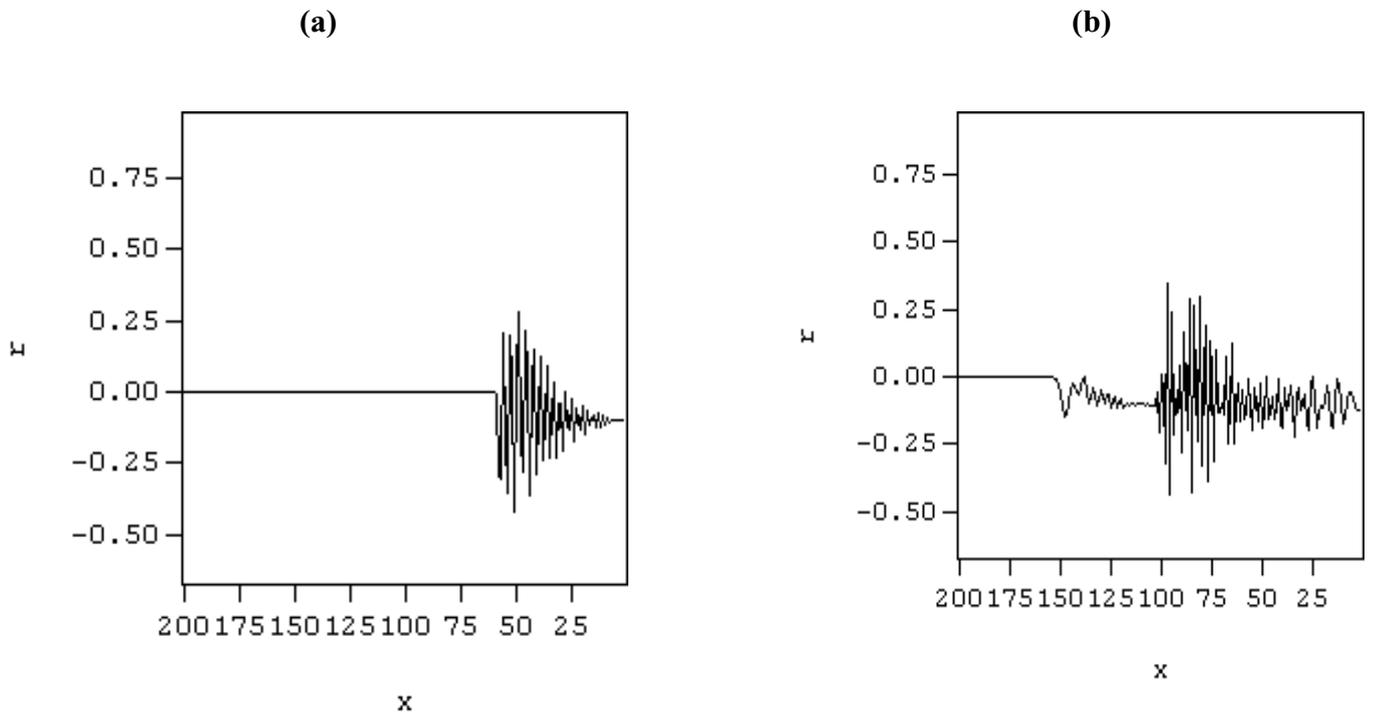

**Figure 6:** Strain profiles for case 4.

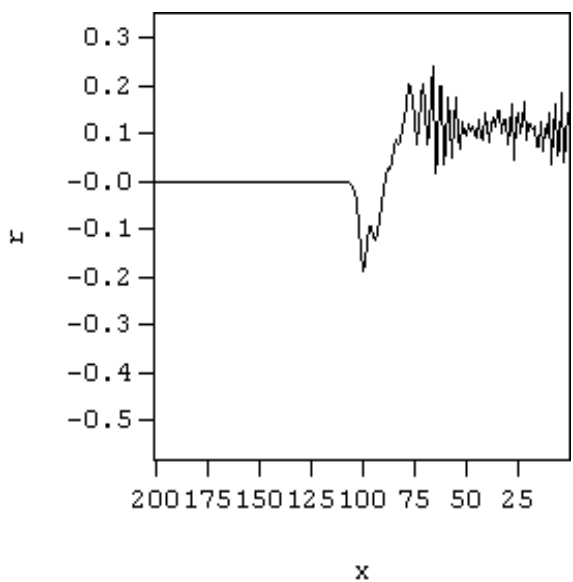

**Figure 7:** Strain profiles for case 5 at a) t=60, b) t=160

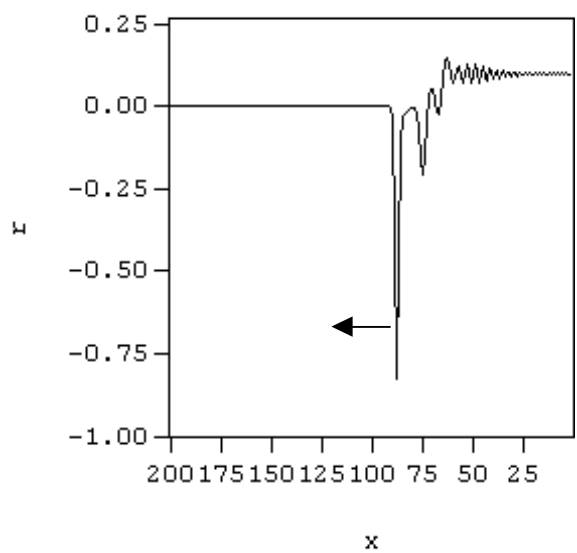 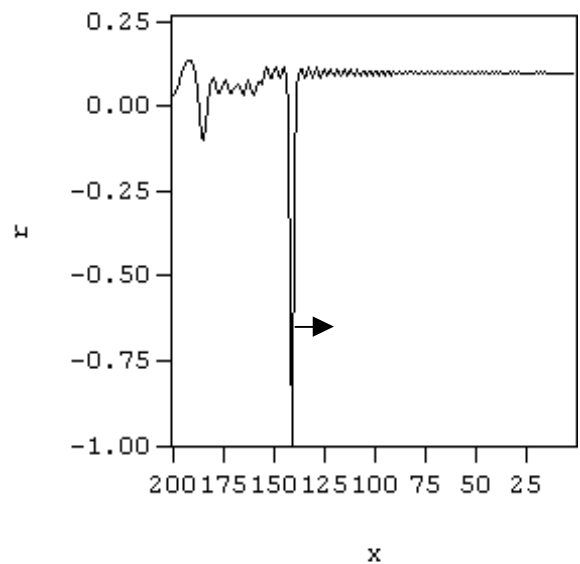

(a)          (b)